\documentclass{aastex62}
\usepackage{amsmath}

\submitjournal{ApJLett}
\shorttitle{Simulations of transient perturbations in an exosphere}
\shortauthors{Leclercq et al.}
\begin{document}

\title{Propagation of Transient Perturbations into a Planet's Exosphere: Molecular Kinetic Simulations}
\correspondingauthor{Ludivine Leclercq}
\email{ll2ad@virginia.edu}

\author{Ludivine Leclercq}
\affil{Material Science and Engineering, University of Virginia, Charlottesville, VA 22903}

\author{Robert E. Johnson}
\affiliation{Material Science and Engineering, University of Virginia, Charlottesville, VA 22903}

\author{Hayley N. Williamson}
\affiliation{Material Science and Engineering, University of Virginia, Charlottesville, VA 22903}

\author{Orenthal J. Tucker}
\affiliation{Goddard Space Flight Center, NASA, Greenbelt, MD 20771}
\begin{abstract}
The upper atmospheres of Mars and Titan, as well as those on many other planetary bodies, exhibit significant density variations vs. altitude that are interpreted as gravity waves. Such data is then used to extract vertical temperature profiles, even when such perturbations propagate through the transition region from a collision dominated regime and into a planet's exosphere. Since the temperature profile is critical for describing the upper atmospheric heating and evolution, we use molecular kinetic simulations to describe transient perturbations in a \textbf{Mars-like upper atmosphere}. We show that the standard methods for extracting the temperature profile can fail dramatically so that molecular kinetic simulations, calibrated to observed density profiles, are needed in this region of a planet's atmosphere.      
\end{abstract}


\section{Introduction} \label{sec:intro}
\indent The physics of the exobase region of the upper atmospheres of exoplanets, solar system planets, planetary satellites, and Kuiper Belt Objects (KBOs) determines their long-term evolution. The behavior of this region has been shown to be sensitive to the molecular composition and the transition from a collision dominated fluid-like regime to a nearly collisionless corona from which escape occurs. Based on MAVEN (Mars Atmosphere and Volatile Evolution) data at Mars and Cassini data at Titan, significant variations in the density structure with altitude are observed from the lower atmosphere into the exosphere. Such perturbations are generally interpreted as gravity waves, which are certainly generated in these atmospheres (e.g., \citealt{snowden_thermal_2013,yigit_high-altitude_2015,england_maven_2017,terada_global_2017}). The density  data is typically analyzed using continuum fluid descriptions of the atmosphere which have been shown to fail well below a planet's exobase (e.g., \citealt{volkov_thermally-driven_2011,volkov_thermal_2013,tucker_thermally_2009,tucker_thermally_2012,tucker_examining_2016,johnson_erratum:_2013,johnson_molecular-kinetic_2013}). On the other hand, molecular kinetic simulations, which are numerical solutions to the Boltzmann equation, can be used to describe the transition from the collision dominated to the nearly collisionless regime giving the thermal structure of the upper atmosphere and the escape rate. Such simulations are especially important as the atmospheric temperature is typically not measured but is extracted from density vs. altitude data assuming local thermodynamic equilibrium. However, we have previously shown that continuum models can fail even when the mean free between collisions is a very small fraction of the scale height \citep{tucker_diffusion_2013,tucker_examining_2016}.\\
\indent  Molecular kinetic simulations have been used extensively to determine the \textit{steady state} behavior of an atmosphere in which the relaxation time scales are short compared to day/night and seasonal time scales. In this paper, the Direct Simulation Monte Carlo (DSMC) method \citep{bird_DSMC_2013} is used to study \textit{transient} events that propagate through the transition region and into the exosphere with emphasis on mass separation and on extraction of the local temperature which is directly calculated in such simulations. The region of interest is a few scale heights below the nominal exobase to a few scale heights above where collisions can be ignored. Perturbations can be produced by transient solar events affecting the absorption of short wavelength radiation, by a heat pulse due to a transient flux of the ambient plasma and pick-up ions, or by a gravity wave formed at depth propagating into this region. \textbf{In this paper we do not try to describe how the observed density perturbations are produced}. Our goal is to better interpret the implications of wave structure observed in the transition region, as this can affect our understanding of the escape rate  and the evolution of an atmosphere \textbf{\citep{walterscheid_wave_2013} }. We first describe the simulations in section \ref{sec:model}. Then, in section \ref{sec:result}, we simulate perturbations in two atmospheres, O only or O+$\mathrm{CO_2}$ using \textbf{Mars-like atmospheric properties, although the results are generally applicable}. Finally, in section \ref{sec:temperature}, we show that the calculated temperature profile differs significantly from the temperatures typically extracted from density variations with altitude indicating molecular kinetic simulations are required to interpret such data. 
\section{Model}\label{sec:model}
\subsection{Description of the DSMC}
\indent In the 1D DSMC method, the motion of atmospheric molecules is followed subject to gravity and mutual collisions using a large number of physical particles each with a statistical weight $w_s$ \cite{bird_DSMC_2013}. Our simulation domain is composed of 55 cells whose sizes range from 6 to 7 km depending on the altitude, with the bottom and top boundaries at 100 and 450 km. These values are subsequently varied to be sure that their choice does not affect the outcome with the cell-sizes chosen to be of order or smaller than local mean free path. The density $n_s$ and the temperature $T_s$ for species $s$ in cell $i$ are computed as:
\begin{equation}\label{eq:density}
n_s(i)=\dfrac{N_s(i)w_s}{V_i}
\end{equation}
\begin{equation}
T_s(i)=\dfrac{m_s}{3k_B}\left(\dfrac{\sum_{p=1}^{N_s(i)} v_p^2w_s}{n_s(i)V_i}-<v_{p_s}(i)>^2\right)
\end{equation}
where $N_s(i)$ is the number of test particles of type $s$ with mass $m_s$ in cell $i$, $k_B$ is the Boltzmann constant, $v_p$ is the velocity of the particle $p$ and $V_i$ is the volume of cell $i$. $<v_{p_s}(i)>$ is the average velocity in the cell $i$ for the species $s$. Particles of species $s$ are assigned a weight $w_s=N_s/N_p$, where $N_s$ is the total column density and $N_p$ is the total number of test-particles created at the initialization. These particles are initially distributed to obtain a barometric density profile with velocities chosen from a Maxwell-Boltzmann (MB) distribution. At each time step, $dt\sim$0.5 s, particles are ejected from the lower boundary using an upward flux, $\Phi_{0_s}=n_{0_s}<v_s>/4$, with $n_{0_s}$ the density and $<v_s>=\sqrt{\tfrac{8k_BT_s}{m_s\pi}}$ the average velocity. Reducing the time step to $dt\sim 0.1$ s did not affect the results shown below. The velocity of the particles entering from the lower boundary is chosen from a Maxwell-Boltzmann Flux distribution \citep{smith_monte_1978}. Particles with energy smaller than the escape energy that cross the upper boundary are assumed to be ballistic. Their trajectories are still computed at each time step until they return to the simulation domain, where they collide with other particles. That is, we track the trajectories of all the particles, even beyond the simulation domain in which we compute the density and temperature. Such particles are often simply reflected, which is adequate when simulating a steady state atmosphere but can fail when simulating \textit{transients}. In these simulations we used a number of cross section estimates but only show results using cross sections from \cite{lewkow_precipitation_2014} recently applied at Mars by \cite{leblanc_origins_2017}. The results are intended to be broadly applicable and can be applied to other atmospheres by scaling (e.g., \citealt{johnson_volatile_2015}).  
\subsection{Simulations parameters}\label{sec:simu_infos}
The effect of perturbations are calculated in either an O or an O+$\mathrm{CO_2}$ atmosphere using gravity and densities like those in Mars upper atmosphere. After the atmosphere reaches steady state, a perturbation is generated by creating a density or a temperature pulse at 150 km of altitude where the atmosphere is collisional. For the simple O atmosphere the density at the lower boundary (100 km) is $10^{10}~\mathrm{cm^{-3}}$, with a temperature of 270 K giving a scale height of $\sim$40 km, an exobase at $\sim$230 km and a mean free path at the lower boundary of $\sim1.5$ km using an average O+O cross section of $4.5\times 10^{-16}~\mathrm{cm^2}$. In the multi-component atmosphere, the density at the lower boundary for O and $\mathrm{CO_2}$ respectively are $1.6\times10^{8}~\mathrm{cm^{-3}}$ and $2.9\times10^{10}~\mathrm{cm^{-3}}$ with a temperature 270 K. Such parameters give a $\mathrm{CO_2}$ scale height of $\sim 15$ km, a $\mathrm{CO_2}$ exobase at $\sim 200$ km, and a mean free path for $\mathrm{CO_2}$ at the lower boundary of $\sim 0.03$ km using an average $\mathrm{CO_2+CO_2}$ cross section $\sim 10^{-14}~\mathrm{cm^2}$. \textbf {Since our goal is to interpret the effect of perturbations that propagate in the transition region, we first generate a relatively large initial density perturbation for a relatively short time}, $\sim$25 s, by adding particles in the cell at 150 km in each time step to maintain a density 2 times the initial local density while maintaining the initial local temperature. \textbf {Reducing this to a more reasonable value did not the change the implications. Therefore, we also initiated a modest} heat pulse, produced by increasing the velocity of particles in the cell at 150 km in each time step to maintain a MB distribution with a temperature $T=300$ K also for $\sim 25$ s.\\
\indent \textbf{As discussed below, both the extreme and modest perturbations exhibit similar behavior as they propagate in the transition region; changes in the amplitude or pulse length gave similar results. To further test this}, we also simulated a wave-like perturbation occurring at the lower boundary of the simulation regime. This was done in the single component atmosphere by varying the incoming flux with time, $t$, at the lower boundary as $\Phi_s(t)=\Phi_{0_s} [1+A\sin\left(B(t-t_0)\right)]$ where $t_0$ is the start time. The quantity $A$ was set to 0.25 which resulted in gravity waves with amplitudes seen frequently in Mars upper atmosphere (e.g., \citealt{yigit_high-altitude_2015,terada_global_2017}). Here $B=\sqrt{-\tfrac{g}{n}\tfrac{dn}{dz}}$ is the Brunt-V\"ais\"al\"a (BV) frequency with $g$ the gravitational acceleration and $(dn/dz)/n$ the inverse of the scale height of the background atmosphere giving a period,  $2\pi/B\simeq 670$ s. Simulations were run varying the surface flux for 5BV periods each with fixed temperature which is equivalent to varying the density or pressure.\\
\section{Perturbation propagation into the exosphere}\label{sec:result}
\begin{figure}[ht]
\centering
\includegraphics[width=16cm]{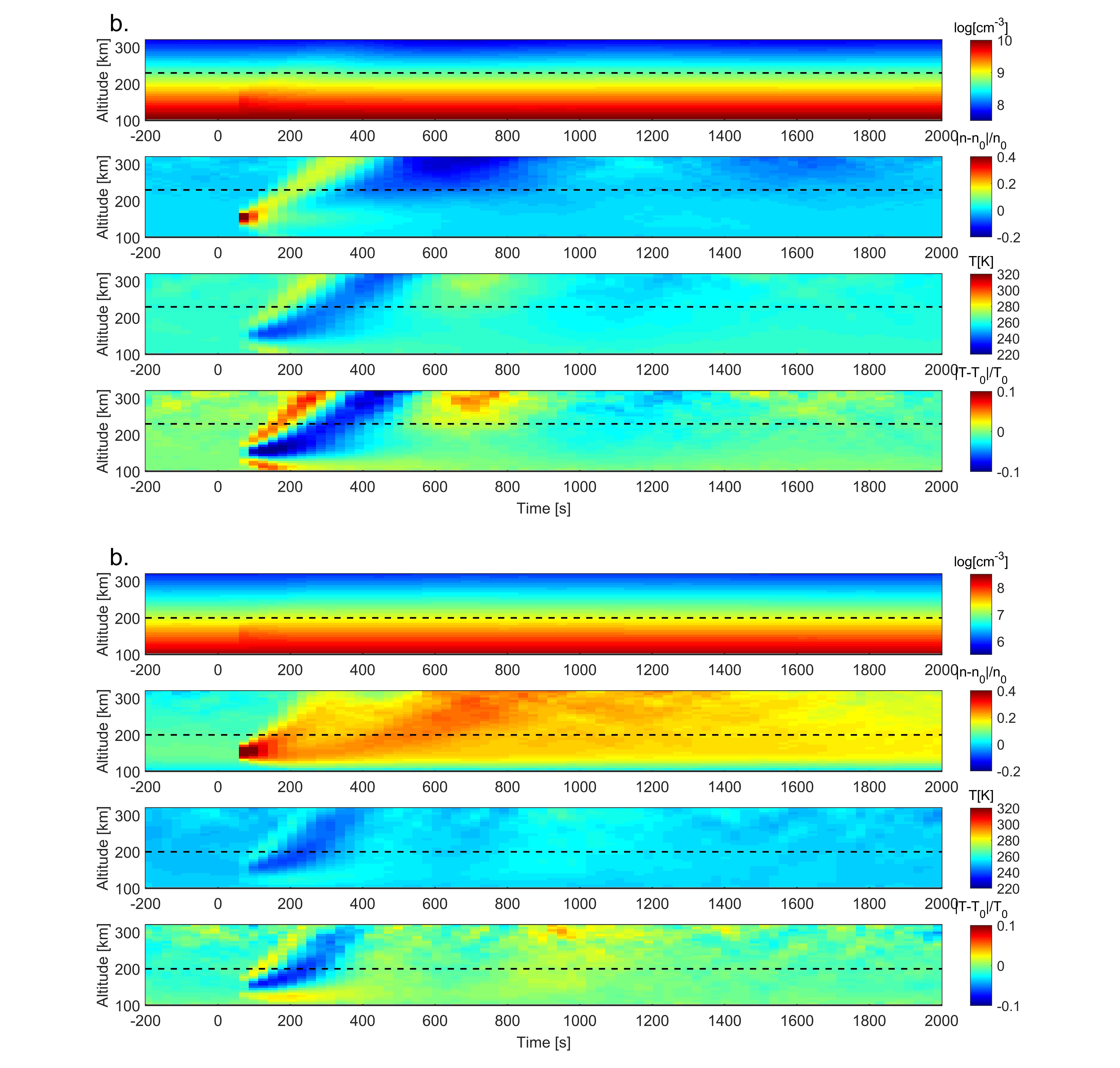}
\caption{\small Evolution of O in an O atmosphere (top panels) and an  O+$\mathrm{CO_2}$ atmosphere (bottom panels) vs. time and altitude following a density pulse $2n_0$ for 50 time steps, $\sim 25$ s. From top to bottom: density in $\mathrm{cm^{-3}}$ ; $(n-n_0)/n_0$  ; temperature in K ; $(T-T_0)/T_0$. Black dotted lines indicate the nominal exobase altitudes described in text.}
\label{fig:O_wave}
\end{figure}
\indent Figure \ref{fig:O_wave} presents the temporal evolution of a \textbf{large initial} density pulse, in an O (top panels) and in an O+$\mathrm{CO_2}$ (bottom panels) atmosphere. The panels (top to bottom) show the density and its amplitude $(n-n_0)/n_0$, and the temperature and its amplitude $(T-T_0)/T_0$, with $n_0$ and $T_0$ the average, steady state values at time 0. The dotted lines indicate the nominal exobase altitudes described above. As the perturbations are produced in the collisional regime, with the mean time between collision short compared to the perturbation time, the collision rate increases as does the pressure so that we find that the speed distribution stays very close to a MB distribution during the perturbation. Therefore, the upward and downward particle flux from the perturbed cell is the MB flux $\Phi(i) \sim n_s(i)<v_i>/4$. Since the faster particles dominate the flow across any boundary, the corresponding energy flux for an MB distribution is $\sim (2k_bT_s)\Phi(i)$ and not $\sim(3k_bT_s/2)\Phi(i)$. Therefore, heat is transiently removed faster than particles following a perturbation, a kinetic effect seen in the simulations when the mean free path between collisions is not negligible. If the mean free path is indeed very small compared to any atmospheric length scale, this difference is equilibrated locally by collisions so that a thermal conductivity can be used. That is not the case in the transition region (e.g., \citealt{tucker_examining_2016}) and it is seen that, even though the very large  perturbation rapidly relaxes, the temperature enhancement precedes the density pulse as discussed further below.\\ 
\indent The pulses (enhancements) propagate upward and downward, locally heating the atmosphere while cooling the perturbed region. At a time $t\sim250$ s, it is seen that the downward propagating pulse appears to be 'reflected'. This feature is only marginally modified by either increasing the height of the perturbation or lowering the boundary. Therefore, the effect is due to the increase in the collision rate in the high density regime below the perturbation and disappears when collisions are suppressed. In a fluid dynamic sense, the perturbation is constrained by the buoyant force in this stable region of the atmosphere. It is also seen that the oxygen density remains larger than the steady state density even after $\sim2000$ s in the multi-component atmosphere unlike in the one-species simulation. In these simulation we chose an O density so that the atoms experience roughly the same number of collisions in the lower atmosphere in both cases. However, collisions of O with the much heavier $\mathrm{CO_2}$ result in a longer residence time in the lower atmosphere, a slower approach to steady state, and a smaller O temperature amplitude. The pulse amplitude, $(n-n_0)/n_0$, continues to grow above the exobase, a feature seen at Mars but becomes suppressed at high altitudes on Mars. \\
\begin{figure}[ht]
\centering
\includegraphics[width=16cm]{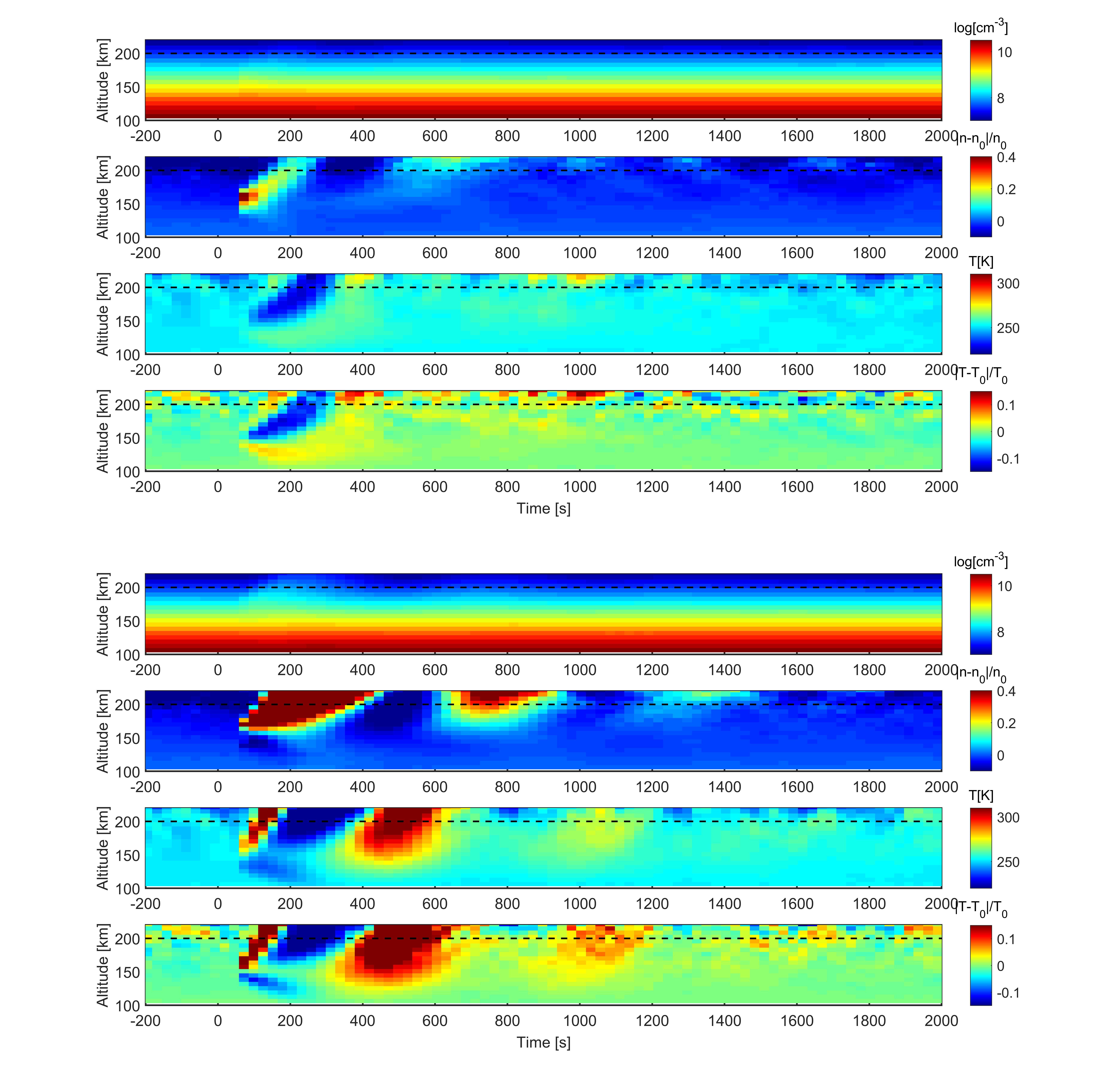}
\caption{\small Evolution of the $\mathrm{CO_2}$ component in an O+$\mathrm{CO_2}$ atmosphere vs. time and  altitude following a ($\sim 25 $ s) pulse at 150 km: top panels, density pulse (2$n_0$); bottom panels, heat pulse ($\Delta T\sim 30$ K). Individual panels as in Figure \ref{fig:O_wave}.}
\label{fig:OCO2_CO2_wave}
\end{figure}
\indent Figure \ref{fig:OCO2_CO2_wave} compares the temporal evolution of $\mathrm{CO_2}$ in an O+$\mathrm{CO_2}$ atmosphere perturbed by the \textbf{large} density pulse (top panels) and a \text{modest} heat pulse (bottom panels). Following the density pulse in the mixed atmosphere, the $\mathrm{CO_2}$ component approaches steady state faster than the O component in Figure \ref{fig:O_wave}. The $\mathrm{CO_2}$ stabilizes faster as they are heavier, have a much larger cross section, and are confined gravitationally to the higher density region. For the same reasons as described for the O atmosphere, the temperature peak precedes the density peak for both temperature and density perturbations.\\
\begin{figure}[ht]
\centering
\includegraphics[width=16cm]{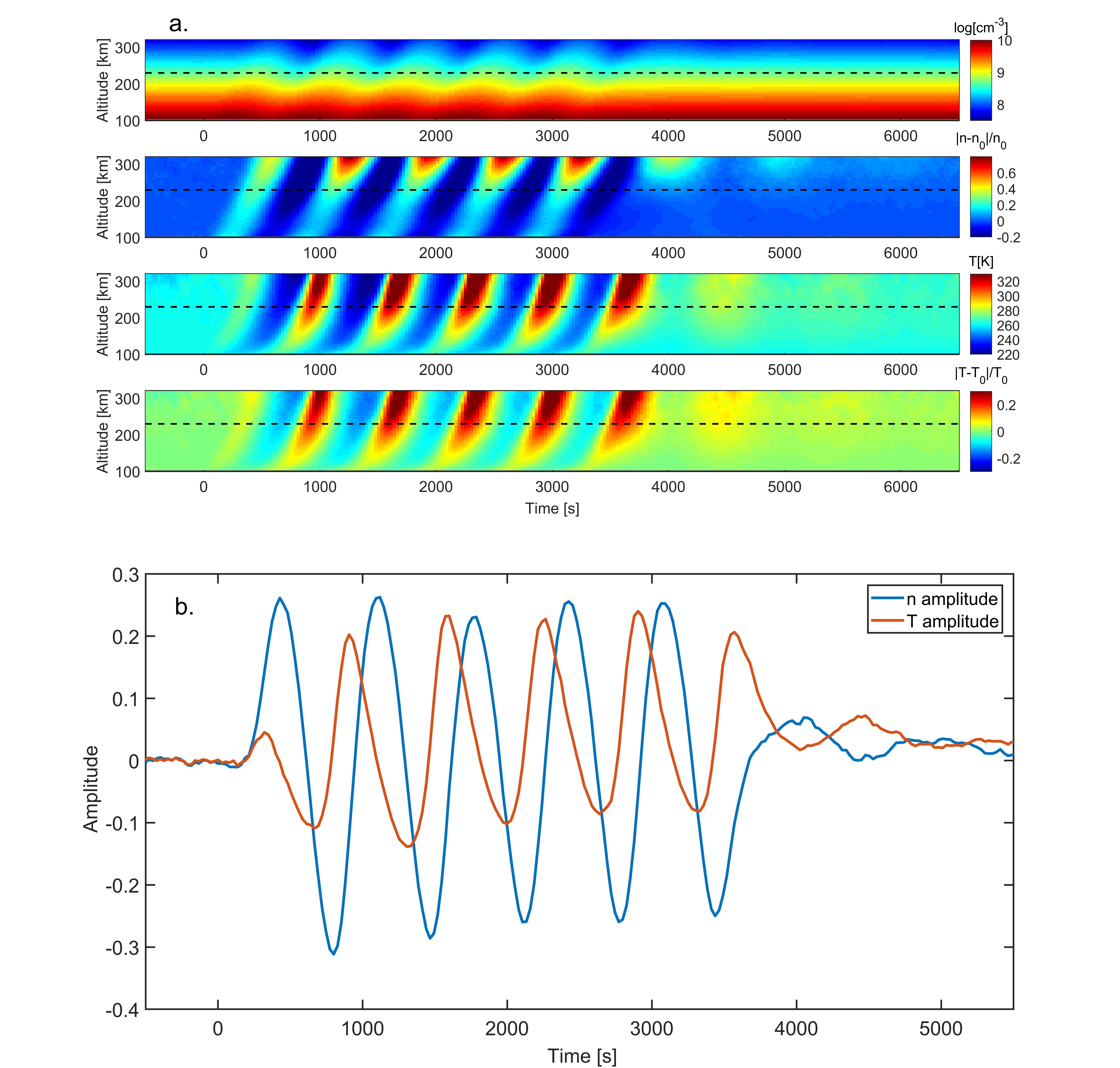}
\caption{\small Response of an O atmosphere in altitude and time: a) injection flux at the lower boundary varies for 5BV periods (period$\sim 660s$) with panels as in Figure \ref{fig:O_wave}. b)$(n-n_0)/n_0$ in blue and $(T-n_0)/T_0$ in red, in function of time, at 230km (exobase): time difference between peaks increase slowly with altitude.}
\label{fig:O_GW_5Bvf}
\end{figure} 
\indent Finally, Figure \ref{fig:O_GW_5Bvf}a) shows the propagation in the transition region of a wave-like perturbation produced at the lower boundary for 5BV periods. As the density at the lower boundary increases and decreases, the collision rate varies affecting the local temperature. It is seen that the wave pattern becomes roughly stable and dies out in $\sim 2$BV periods, which is $\sim 20$ minutes \textbf{in this model atmosphere}. Figure \ref{fig:O_GW_5Bvf}a) shows the temporal evolution of the density and temperature amplitudes with altitude. The wave amplitudes are seen to increase with altitude as expected. We find again that the temperature pulse precedes the density pulse at all altitudes due to the more rapid transport of thermal energy in this region of the atmosphere. This result \textbf{is consistent with the large amplitude density pulse and more modest temperature pulse and} is seen explicitly in Figure \ref{fig:O_GW_5Bvf}b) at the exobase altitude. We also find, not surprisingly, that the time separation between peaks grows slowly with altitude as the mean free path between collisions increases.\\
\section{Temperature extraction}\label{sec:temperature}
\indent In the extensive analysis of the upper atmosphere of Mars (\citealt{yigit_high-altitude_2015,england_maven_2017,liu_longitudinal_2017,walterscheid_wave_2013}) the observed variations in the vertical structure of the density vs. altitude, interpreted as gravity waves, were used to extract the temperature structure following the 1D method used by \citealt{snowden_thermal_2013} for Titan's upper atmosphere. The hydrostatic law was used to calculate a pressure vs. altitude profile from smoothed density data measured by the Neutral Gas and Ion Mass Spectrometer (NGIMS) on MAVEN. Based on the ideal gas law, that profile was subsequently used to extract the local temperature vs.altitude (\cite{england_maven_2017,liu_longitudinal_2017}). As the perturbations appeared to propagate into the region above the nominal exobase ($\sim$200 km) \citealt{yigit_high-altitude_2015} and others cautioned the method could be problematic. Using the results in Figures \ref{fig:O_wave} to \ref{fig:O_GW_5Bvf} we show that not only were these cautionary remarks correct but the extracted temperature profiles can be incorrect. \\
\indent The integration of pressure vs. altitude from the measured density data requires a value for the pressure, $P_u$, at the upper limit of the data, $n_u$.  Assuming $P_u=n_uk_BT_u$, the temperature at the upper boundary, $T_u$, is estimated using:
\begin{equation}
\dfrac{d\log n(r)}{dr}=\dfrac{mg(r)}{k_BT_u}\left(\dfrac{\alpha}{C_p}- 1\right)
\end{equation}
where $C_p$ is the specific heat, $g(r)$ is the gravitational acceleration, $n$ the density and $r$ the distance to the center of the body (\cite{snowden_thermal_2013}). Although $\alpha$ was varied from 0 to $\pm0.5$ to take into account uncertainties in the extrapolation, we only show profiles using $\alpha=0$. Changing alpha changes the temperature values at high altitudes but does not improve the agreement with the simulations.\\
\begin{figure}[ht]
\centering
\includegraphics[width=13cm]{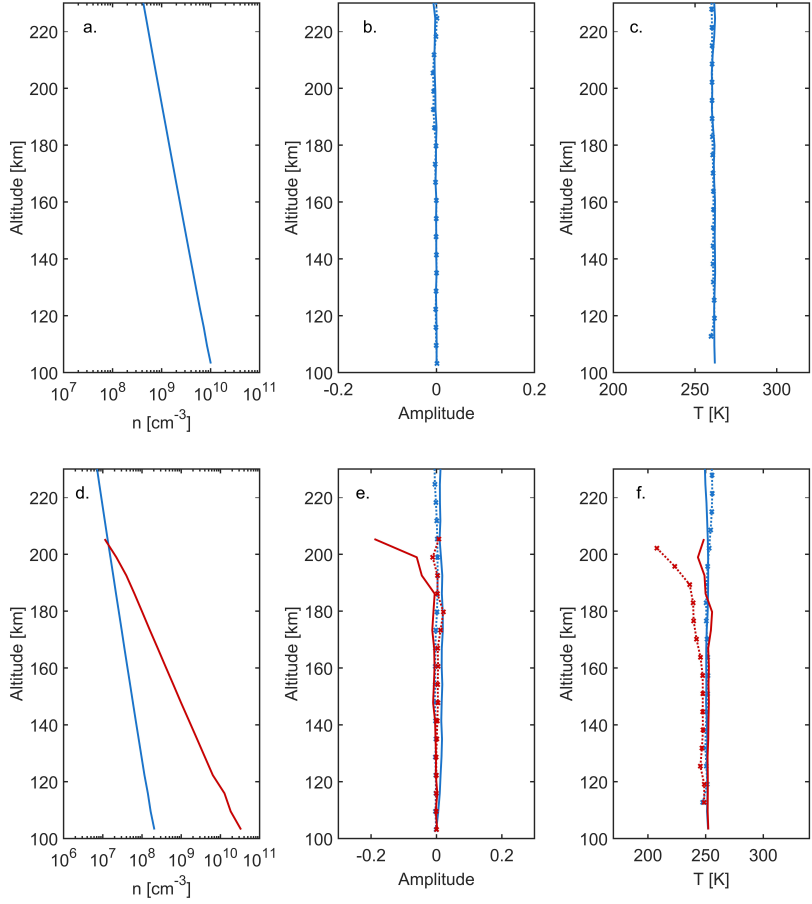}
\caption{\small Steady state: top 3 panels O atmosphere; bottom 3 panels $\mathrm{O+CO_2}$, O (blue), $\mathrm{CO_2}$ (red): a),d) Simulated density; b),e) $(n-n_0)/n_0$ (solid), $(T-T_0)/T-0$ (dotted). c),f) Simulated temperature (solid), extracted temperature (dotted). Weight differences between O and $\mathrm{CO_2}$ account for differences in statistics above 180 km.} 
\label{fig:snowden_st}
\end{figure}
\indent Figure \ref{fig:snowden_st} shows the steady state density and temperature (solid lines) from our O and O+$\mathrm{CO_2}$ simulations. Figure \ref{fig:snowden_st}b) and e) confirm that the density and temperature amplitudes at steady state are nearly zero and the extracted and simulated temperatures are in agreement to within the uncertainties. In the following, the DSMC simulated kinetic temperature is compared to the temperature extracted from the calculated density profile in Eq. \ref{eq:density} which are in rough agreement in steady state (solid and dotted lines in \ref{fig:snowden_st}c) and f)). In the discussions below, only density values below the nominal exobase altitudes are compared.\\
\begin{figure}[ht]
\centering
\includegraphics[width=13cm]{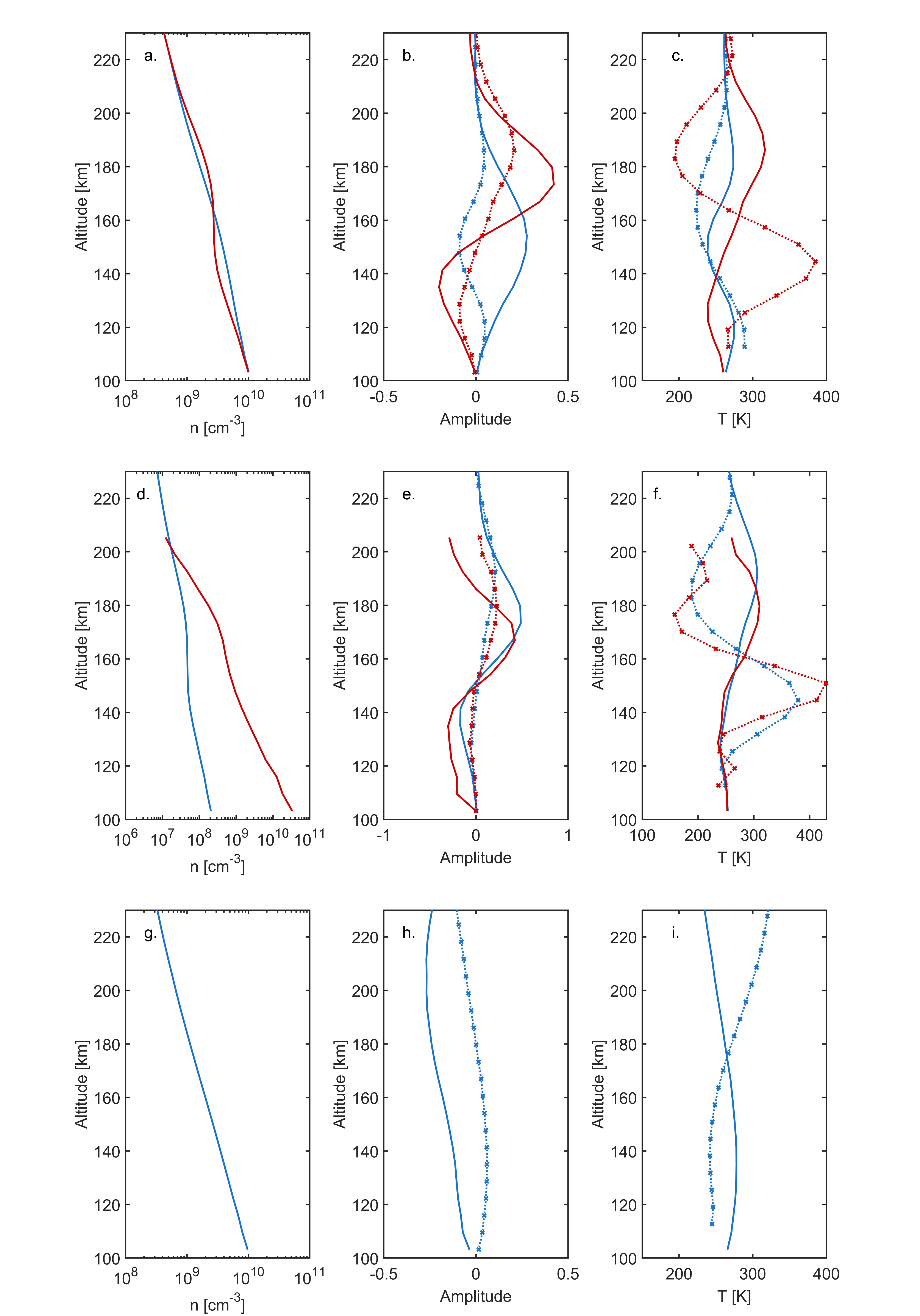}
\caption{\small a)b)c) O atmosphere. Results obtained  $\sim 55$ s after a density pulse (blue) and heat pulse (red). d)e)f) results obtained $\sim 55$ s after a heat pulse in a O (blue) + $\mathrm{CO_2}$ (red) atmosphere. g)h)i) 1 BV wave perturbation in an O atmosphere $\sim 720$ s after initiation. a)d)g) Density ($\mathrm{cm^{-3}}$); b)e)h) n and T amplitudes (solid and dotted); c)f)i) simulated (solid) and extracted (dotted) temperatures in K.}
\label{fig:snowden_comparison}
\end{figure}
Figure \ref{fig:snowden_comparison} shows our key results. The vertical profiles at 60s after the perturbation in Figures \ref{fig:O_wave} and \ref{fig:OCO2_CO2_wave} (top panels: pulse in the O atmosphere: density in blue, temperature in red ; middle panels: temperature pulse in the O+$\mathrm{CO_2}$ atmosphere; bottom panels: for wave perturbations, from Figure 3, $\sim 720$ s after the perturbation starts, the first propagating pulse). From the left to the right, the panels give the density, the amplitude, and the extracted and simulated temperatures profiles. For the density pulse, the thermal wave is seen to precede the density wave causing a transient thermal depression in the perturbed region, with thermal peaks propagating away from the region as discussed above. In contrast to this, the extracted temperature simply follows the form of the pressure wave gradient. This results in a difference of $\Delta T\sim-90$ K with respect to the steady state atmosphere, overestimating the local cooling of the atmosphere ($\Delta T\sim-60$ K). When the perturbation is due to a heat pulse (red curves), the kinetic and extracted temperature profiles are almost out of phase. In particular, between 180 km and 220 km the model shows the atmosphere is heated by the perturbation with a temperature increase of $\Delta T \sim 50$ K while the extracted temperature shows it cooled ($\Delta T\sim -70$ K). These temperatures are also in serious disagreement in the O+$\mathrm{CO_2}$ atmosphere (middle panels: O in blue, $\mathrm{CO_2}$ in red). Below 160 km the simulations predict a small thermal perturbation whereas the extracted T reaches about 340 K for each species, which would require heating by $\Delta T\sim 70$ K. Above 160 km, the simulated temperatures peak at $\sim10$ K for each species. The extracted temperature on the other hand requires local cooling of $\Delta T\sim-50$ K for the O and $\sim-100$ K for the $\mathrm{CO_2}$ component. Finally, for a wave like perturbation propagating into this region from the lower atmosphere (bottom panels), the extracted and simulated temperatures, in the bottom right hand panel, not only disagree but are out of phase. Therefore, the published thermal profiles in the upper atmospheres of Mars and Titan that are extracted from the density profiles need to be re-examined based on a molecular kinetic model.
\section{Summary}\label{Summary}
Molecular kinetic simulations were carried out to describe how \textbf{extreme and modest} atmospheric perturbations propagate into and through the transition region for both single component and two component atmospheres. Because the mean free path between collisions is not negligible, the temperature pulse is out of phase with the density pulse, unlike what is assumed in 1D continuum extraction models typically used. Not surprisingly, the heavy species quench faster than the light species, and we found that, although the density amplitude grows as the perturbations propagate upward through the transition region, the growth with altitude differs considerably from what is expected from a linear theory (e.g.,\citealt{hines_internal_1960,oberheide_tides_2015}). Well above the nominal exobase, the amplitudes at Mars, for instance, eventually decrease requiring 2D, multicomponent simulations which are now being carried out. Finally, and most important, published temperature profiles extracted from measured density profiles \textit{below} the exobase, but in the transition region, are likely incorrect. We show this is the case even when the observed density variations are driven by wave-like perturbations from below. These results are generally applicable by scaling but indicate that molecular kinetic simulations are needed to correctly interpret the wave-like features in the transition region of a planet's atmosphere.\\
\acknowledgments The authors would like to thanks D. Snowden for help on the extraction method and C. Schmidt and A. Volkov for helpful comments. Work supported by the Cassini mission through SwRI and a NASA Planetary Data Grant NNX15AN38G.\\



\end{document}